\documentclass{Rinton-P9x6}
\usepackage{amsbsy,latexsym}
\usepackage{amsfonts}
\usepackage{amssymb}
\usepackage[mathscr]{eucal}

\newcommand{\vn}{\vec{n}}
\newcommand{\vM}{\vec{M}}
\newcommand{\vMx}{\vec{M}(x)}
\newcommand{\vVx}{\vec{V}(x)}

\newcommand{\kn}{\ket{\vec{n}}}
\newcommand{\kMx}{\ket{\vec{M(x)}}}
\newcommand{\eqref}[1]{(\ref{#1})}

\newcommand{\la}{\lambda}
\newcommand{\e}[1]{\exp\left(#1 \right)}

\newcommand{\ket}[1]{\vert #1 \rangle}
\newcommand{\bra}[1]{\langle #1 \vert}

\newcommand{\NN}{{\mathscr N}}

\begin{document}

\title{Optimal reconstruction of a pure qubit state with local measurements}

\author{E.~Bagan, M.~Baig, A.~Monras and  \underline{R.~Mu{\~n}oz-Tapia} }

\address{Grup de F{\'\i}sica Te{\`o}rica \& IFAE, Facultat de Ci{\`e}ncies,
Edifici Cn, Universitat Aut{\`o}noma de Barcelona, 08193 Bellaterra
(Barcelona) Spain }


\maketitle

\abstracts{ We analyse the reconstruction of an unknown pure qubit
state. We derive the optimal guess that can be inferred from any
set of measurements on $N$ identical  copies of the system with
the fidelity as a figure of merit. We study in detail the
estimation process with individual von Neumann measurements and
demonstrate that they are very competitive as compared to
(complicated) collective measurements. We compute the expressions
of the fidelity for large $N$  and show that individual
measurement schemes can perform optimally in the asymptotic
regime.}

\section{Introduction}
State estimation is a fundamental issue in Quantum Information
from both theoretical and practical points of view. Imagine we are
asked to reconstruct the \emph{unknown} state of a quantum system.
This can only be done by  performing  measurements on  an ensemble
of identically prepared systems.  With an infinite ensemble of
copies, the state could be determined exactly. In practice,
however, we have access to a limited number of copies and the
state can only be determined approximately.\cite{holevo} In this
context, three essential questions arise: \textit{i}) what is the
optimal measurement scheme?, \textit{ii})  what is the best
reconstructed state?, and \textit{iii}) how good is the overall
estimation process?.

In recent years a lot of work has been devoted to answer these
questions for different
settings.\cite{mp,derka,pure-states,us-product,density} The
optimal strategies, which provide the ultimate limits that can be
achieved, have been identified in several interesting cases.
However, they usually involve collective measurements (CM), i.e.~a
generalised measurement on all copies at the same time. These,
although very interesting from the theoretical point of view, are
very difficult to implement in practice. Far more interesting for
experimentalists are individual von Neumann measurements for they
can be readily performed in a laboratory.  In this case however,
fewer analytical results are
known.\cite{locals,us-local,us-density} Here, we present some
theoretical results along these lines.

We focus on the problem of estimating the most basic quantum
state, a pure qubit, with physically realizable von Neumann
measurements. We study quantitatively tomographic inspired
schemes,\cite{tomography,likelihood} but also consider the most
general individual measurement procedure, i.e. when depending on
the previous outcomes, one allows to optimally adapt  the
measurement on the subsequent copy.\cite{us-local} To ease our
presentation, we will loosely write \textit{local measurements}
for individual von Neumann measurements. Our aim is to investigate
how good these local measurements are as compared to the
collective ones. We use the fidelity as the figure of merit
quantifying the quality of the estimation process and compute the
analytical expressions of the average fidelity for large $N$. Two
interesting situations will be analysed  which we will refer to as
2D and 3D. In 2D the qubit is only known to be a state of the
equator of  the Bloch sphere. This is formally equivalent to phase
estimation. In 3D no prior knowledge of the qubit is assumed.

This paper is organised as follows. In next section we obtain the
optimal state that can be inferred from any set of measurements.
This is a  general result valid for any measurement scheme (either
collective or local) and for any a priori probability
distribution. The 2D and 3D case with fixed (non-adaptive) local
measurements is studied in section~\ref{fixed local}. The most
general local scheme is presented in section~\ref{optimal local}.
We conclude with a summary and outlook for further work.

\section{Optimal guess}\label{optimal guess}
The estimation procedure goes as follows. Assume that we are given
an ensemble of $N$ copies of the qubit state, which we denote by
$\kn$, where $\vn$ is the unique unit vector on the Bloch sphere
that satisfies $\kn\bra{\vn}=(1+\vec{n}\cdot \vec{\sigma})/2$ and
$\vec{\sigma}$ are the usual Pauli matrices. After performing a
set of measurements on the $N$ copies of the qubit,  one obtains a
set of outcomes symbolically denoted by $x$. Based on $x$, an
estimate for $\kn$ is guessed, $\kMx$. How well $\kMx$
approximates  the signal state $\kn$ is quantified by the
fidelity, defined as the overlap
\begin{equation}\label{f}
  f_n(x)\equiv|\bra{\vn}\vMx \rangle|^2={1+\vn \cdot\vMx \over
  2}.
\end{equation}
Eq.~(\ref{f}) is a kind of ``score": we obtain `1' for a perfect
determination ($\vec{M}=\vec{n}$) and `0' for a completely wrong
guess ($\vec{M}=-\vec{n}$). Our aim is to maximize the average
fidelity, hereafter fidelity in short, over the initial
probability and all possible outcomes,
\begin{equation}\label{f-1}
    F\equiv\langle f \rangle =
\sum_{x} \int dn\, f_n(x)\;p_{n}(x),
\end{equation}
where $p_{n}(x)$ is the probability of getting outcome $x$ if the
signal state was $\ket{\vec{n}}$, and  $dn$ is the a priori
probability distribution. For a completely unknown qubit, $dn$ is
the invariant measure on the two-sphere  (on the unit circle in
2D). Eqs.~(\ref{f}) and (\ref{f-1}) can be rewritten as
$F=\frac{1}{2}[1+\sum_x \vVx \cdot \vMx ]$, where
\begin{equation}\label{v-optimal}
    \vVx=\int dn \, \vn\, p_n(x).
\end{equation}
It is obvious that the choice
\begin{equation}\label{m-optimal}
    \vMx=\frac{\vVx}{|\vVx|}
\end{equation} maximizes the value of $F$, which then
reads
\begin{equation}
    F=\frac{1}{2}\left(1+\sum_{x}|\vVx|\right).
    \label{f-optimal}
\end{equation}
Eq.~\eqref{m-optimal} gives the best inferred state and
Eq.~\eqref{f-optimal} the maximum fidelity that can be obtained
for \emph{any} a priori probability and \emph{any} measurement
scheme specified by the conditional probabilites $p_{n}(x)$.

In the next sections we show how these simple results can be used
to improve the estimation procedure. From now on we will only
consider the 2D and 3D isotropic probability distributions given
by $dn=d\phi/(2\pi)$ and $dn=\sin\theta d\theta d\phi/(4\pi)$,
respectively. A fidelity with no explicit label refers to the 3D
case.

\section{Fixed local  measurements}\label{fixed local}

Before dealing with local measurements, let us recall some known
results of the collective schemes. The optimal fidelity for the 2D
case is\cite{derka}
\begin{equation}\label{f-2d}
    F^{\rm (2D)}_{\rm CM}=\frac{1}{2}+\frac{1}{2^{N+1}} \sum_i \sqrt{\pmatrix{N \cr i}\pmatrix{N \cr
    i+1}} \stackrel{\scriptsize N\to \infty}{\to} 1-\frac{1}{4 N}+
    \cdots,
\end{equation}
whereas for the 3D case reads\cite{mp}
\begin{equation}\label{f-3d}
    F_{\rm CM}=\frac{N+1}{N+2}\stackrel{\scriptsize N\to \infty}{\to} 1-\frac{1}{ N}+
    \cdots,
\end{equation}
These results could, in principle,  also be derived
from~\eqref{f-optimal}. Notice that $ F^{\rm (2D)}_{\rm CM}>F_{\rm
CM}\  \forall N$, as it should, since in the 2D case we have more
a priori information about the state than in the 3D case. These
results are the absolute upper bound for {\it any} measurement
scheme.

Let us now turn our attention to local measurements. Any
individual von Neumann measurement is represented by two
projectors $O(\pm \vec{m})=(1\pm \vec{m}\cdot \vec{\sigma})/2$,
where $\vec{m}$ is a unit Bloch vector characterizing the
measurement (in a spin system, e.g., $\vec{m}$ is the orientation
of a Stern-Gerlach). Quantum state tomography tells us that, given
a large number of copies, von Neumann measurements along two
(three) fixed orthogonal directions, $x,y,(z)$, are sufficient to
reconstruct the state.

Consider $N=2\NN$ ($3\NN$) copies of the state $\kn$. After $\NN$
measurements in each axis, we obtain a set of outcomes $+1$ and
$-1$ with relative frequencies $\alpha_i$ and $1-\alpha_i$,
respectively. This occurs with probability
\begin{equation}\label{probability}
  p(\alpha|\vn)=  \prod_{i=x,y,(z)}\left(\begin{array}{c}
                              \NN\\
                              \NN \alpha_i
                       \end{array}
                  \right)\left(\frac{1+n_i}{2}\right)^{\NN\alpha_i}
                  \left(\frac{1-n_i}{2}\right)^{\NN(1-\alpha_i)},
\end{equation}
where  $n_i$ are the projections of the vector $\vn$ in each
direction and we have used the shorthand notation
$\alpha=\{\alpha_i\}$. Since the expectation value of
$\vec{\sigma}$ is $\bra{\vn}\vec{\sigma}\kn=\vec{n}$ one is driven
to propose a guess
\begin{equation}\label{cl-guess}
  M_{\rm T\,i}(\alpha)=\frac{
  2\alpha_i-1}{\sqrt{\sum_j (2\alpha_j-1)^2}},
\end{equation}
where the subscript stands for tomographic. Notice the presence of
a normalization factor such that $|\vM_{\rm T}|=1$, therefore
$\vM_{\rm T}$ always corresponds to a physical pure state.
Actually, \eqref{cl-guess} is the guess for pure states of maximum
likelihood procedures.\cite{likelihood} The law of large numbers
ensures that $\vec M_{\rm T}\stackrel{\NN\to
\infty}{\longrightarrow}\bra{\vn}\vec{\sigma}\kn=\vn$, but our
main goal is to know the rate at which this limit is attained.

 The asymptotic fidelity can essentially be computed by means
 of the  following
\emph{systematic} approximations (see Bagan {\it et
al.}\cite{us-product,us-local,us-prep} for more details). First,
use the central limit approximation in~\eqref{probability}
\begin{equation}
 \pmatrix{\NN\cr\al \NN} q^{\al \NN} (1-q)^{(1-\al)\NN}\to
    {1\over\sqrt{2\pi \NN q(1-q)}}
\e{-{\NN\over 2}{(\al-q)^2\over q(1-q)}}+\cdots
    \label{stirling}
\end{equation}
with $q=(1+n_{j})/2$. Second, transform the discrete sum in into
an integral using the Euler-McLaurin formula  
\begin{equation}
 \sum_{j=1}^{\NN} {1\over \NN}f(j/\NN) =\int_0^1 dx f(x)+
{f(1)-f(0)\over 2\NN} +{f'(1)-f'(0)\over12\NN^2}- \cdots .
\label{euler-mac}
\end{equation}
The change of variables $r_i= 2\alpha_i-1$, suggested
by~\eqref{cl-guess}, proves to be useful to simplify the
expressions. Finally,
 use saddle point techniques to evaluate the integrals. This just
amounts to consider that the value  of the integrals  is dominated
by the minimum of the exponent in \eqref{stirling} and to expand
systematically around this point.

\subsection{2D results.}
For the tomographic guess~(\ref{cl-guess}), and using the
techniques described above, we obtain the following asymptotic
expression of the fidelity~\eqref{f-1}:
\begin{equation}
    F^{\rm (2D)}_{\rm T}=1-{3\over 8}{1\over N}+\dots.
    \label{f-2d-cl}
\end{equation}
Note that $F_{\rm T}^{\rm (2D)}$ approaches unity linearly in
$1/N$. In this sense, one may argue that the tomographic approach
is qualitatively similar to the optimal collective scheme.
Nevertheless, the coefficient of the first correction is  a 50\%
larger than the optimal one \eqref{f-2d}.

{}From our discussion in section~\ref{optimal guess}, we know that
there is a better guess that can be inferred from the \emph{same}
set of measurements. It is given by~\eqref{m-optimal}, with the
outcomes  labelled by $x=\{\al_x,\al_y\}$ and the probabilities
again given by~\eqref{probability}. The fidelity is then $F_{\rm
OG}=1/2(1+\sum_{\al}|\vec{V}(\al)|)$.  The analytical calculation
of the large $N$ limit is now more involved, mainly due to the
presence of the modulus, but can be performed basically with  the
same techniques.\cite{us-local} It reads
\begin{equation}
    F^{\rm (2D)}_{\rm OG}=1-{1\over4}{1\over N}+\cdots,
    \label{f-2d-og}
\end{equation}
where OG stands for optimal guess. This is a remarkable result.
Provided the optimal guess is used, the most basic estimation
strategy, namely with local and minimal fixed von Neumann
measurements, saturates asymptotically the optimal CM
bound~\eqref{f-2d}.
\bigskip

\subsection{3D results}
The same analysis can be carried out in the 3D case, i.e. when
$\kn$ is a completely unknown qubit pure state. The calculations
are rather more difficult, but can be done analytically till the
end. For the tomographic guess~\eqref{cl-guess} we obtain
\begin{equation}
    F_{\rm T}=1-{6\over 5}{1\over N}+\dots,
    \label{f-3d-cl}
\end{equation}
whereas for the optimal guess
\begin{equation}
    F_{\rm OG}=1-{13\over 12}{1\over N}+\dots.
    \label{f-3d-og}
\end{equation}
As expected,  $F_{\rm OG}>F_{\rm T}$. Notice that, again,  the
first correction of the fidelity goes linearly with $1/N$, now
with a coefficient very close to one.  However, in contrast to the
2D case, the improvement of the optimal guess is not sufficient to
saturate the CM bound for which $F=1-1/N+\cdots$.

\section{Optimal local measurements}\label{optimal local}

The local measurements discussed so far were the most basic ones:
fixed and minimal.  We have not considered yet  local schemes in
full. In particular, classical communication, i.e. the possibility
to adapt the orientation of the measuring devices depending on
previous outcomes, was not exploited. In this section, we obtain
the optimal scheme in this general setting and show explicit
results for low $N$. For large $N$, we also obtain the asymptotic
expression of the fidelity. Hereafter only the general case 3D
will be considered.

We need first to introduce a suitable notation to include
arbitrary orientations of the devices and classical communication.
Consider the set of von Neumann measurements specified by the
collection of Bloch vectors $\{\vec{m}_k\}$.  The set of outcomes
$x$ can be expressed as an $N$-digit binary number
$x=i_{N}i_{N-1}\cdots i_{2}i_{1}$, where $i_k$ ($=0,1$). The most
general local measurement is realized when we allow
$\vec{m}_{k+1}$ to depend also on the list of previous outcomes
$i_{k} i_{k-1}\cdots i_{2}i_{1}\equiv x_{k}$ (hence, $x=x_N$). We
thus write $\vec{m}(x_k)$ instead of $\vec{m}_k$. Note that $\vec
m(x_{k})$ must satisfy the von Neumann condition
\begin{equation}\label{von-neumann}
  \vec m(1x_{k-1})=-\vec m(0x_{k-1}).
\end{equation}
For any set of outcomes, the optimal guess is given
by~\eqref{m-optimal} and~\eqref{v-optimal}, where now the
conditional probability is
\begin{equation}\label{probability-general}
   p_n(x)=\prod_{k=1}^{N}{1+\vec n\cdot\vec m(x_{k})\over2}
\end{equation}
and the fidelity reads
\begin{equation}
    F=\frac{1}{2}\left( 1+ \sum_{x=00\cdots0}^{2^{N}-1}\left|
    \int dn\, \vec n\ \prod_{k=1}^{N}{1+\vec n\cdot\vec
    m(x_{k})\over2}\right|\right).
    \label{f-og-general}
\end{equation}
The optimal  scheme is the one that maximizes~(\ref{f-og-general})
over a set of vectors $\{ \vec{m}(x_k) \}$ with the von Neumann
constraint~\eqref{von-neumann}.

\subsection{Low $N$ cases}

$N=2$. Here, there are three independent Bloch vectors vectors:
$\vec{m}(0)$, $\vec{m}(00)$ and $\vec{m}(01)$ (the other three are
obtained using Eq.~\ref{von-neumann}). The first vector
$\vec{m}(0)$ is arbitrary and can be fixed at will. The optimal
fidelity is then obtained by maximizing (\ref{f-og-general}) with
respect to $\vec{m}(00)$ and $\vec{m}(01)$. A straightforward
calculation yields the following conditions:
\mbox{$\vec{m}(0)\cdot \vec{m}(00)=0=\vec{m}(0)\cdot\vec{m}(01)$}.
Note that
 $\vec{m}(00)$ and $\vec{m}(01)$
do not need to be equal, they are only required to be orthogonal
to $\vec{m}(0)$. Substituting back in~\eqref{f-og-general} one
finds  $F^{(2)}=(3+\sqrt{2})/6$. This is the largest value of
fidelity that can be obtained with two copies  and local
measurements. Obviously $(3+\sqrt{2})/6< F_{\rm CM}=3/4$. The
optimal guess is easily obtained from Eq.~\ref{m-optimal},
$\vec{M}^{(2)}(x)=[\vec{m}(x_2)+\vec{m}(x_1)]/\sqrt{2}$. This is a
very gratifying result: $\vec{M}^{(2)}(x)$ is the `weighted' sum
of the outcomes.

 The case $N=3$ is very similar. The optimal Bloch
vectors, $\vec{m}(x_1)$, $\vec{m}(x_2)$, $\vec{m}(x_3)$, are found
to be mutually orthogonal. They can be  chosen to coincide with
three fixed (i.e. independent of $x$) directions. Thus for  $N=3$
(as well as for $N=2$) the optimal estimation schemes based on
local measurements do not require classical communication. For
each outcome $x$ the optimal guess is $\vec
M^{(3)}(x)=[\vec{m}(x_3)+\vec{m}(x_2)+\vec{m}(x_1)]/\sqrt{3}$,
which is a straightforward generalization of~$\vec{M}^{(2)}$. The
fidelity is $F^{(3)}=(3+\sqrt{3})/6$. These  results could somehow
be anticipated: if $O(\vec m)\kn \not=0$ we can only be sure that
 $\vn \neq -\vec{m}$.  It is then reasonable to explore the plane orthogonal to
$\vec m$ with the next copy of $\kn$. Thus, the optimal Bloch
vectors $\vec m(x_{k})$ tend to be mutually orthogonal.

 The case $N=4$ is more complex, since four mutually
orthogonal vectors cannot fit onto the Bloch sphere.
 We do not reproduce here the explicit expressions of the optimal
vectors\cite{us-local,us-prep}. Instead we would like to point out
some properties of the solution which, in turn, will bring us
insight as to how to compute the asymptotic limit. We observe that
the optimal Bloch vectors now depend on the outcomes of the
previous measurements. Therefore classical communication does play
a crucial role for $N>3$. One also sees that the third measurement
probes the plane orthogonal to the vector one would guess from the
first two outcomes and analogously does the fourth measurement.
The fidelity in this case reads $F^{(4)}=0.8206$, which is just
$1.5\%$ lower than the absolute CM bound $F_{\rm
CM}^{(4)}=5/6=0.8333$. The maximal fidelities for $N=5,6$ are
$F^{(5)}=0.8450$ and $F^{(6)}=0.8637$.

\subsection{Asymptotic fidelity}

We can  finally compute the asymptotic fidelity of the optimal
local scheme. Suppose we have performed a (large) number $N_{0}$
of measurements and obtained an  optimal guess $\vec M_{0}$. It is
clear that the subsequent guesses will hardly differ from $\vec
M_{0}$. It is also clear from our results of low $N$ that the
following measurements will basically probe the orthogonal plane
of $\vec{M}_0$. Hence, a good approximation to the optimal local
strategy would be to consider: \emph{a}) fixed measurements in the
orthogonal plane to $\vM_0$ (i.e. along two orthonormal vectors
$\vec u$, $\vec v$ of the plane) and  \emph{b}) a guess of the
form $\vec M(x)\approx \vec M_{0}\cos\omega+(\vec u \cos\tau+\vec
v \sin\tau)\sin\omega$, where  $\omega =\la \sqrt{ (2
\alpha_{u}-1)^2+(2 \alpha_{v}-1)^2}$, $\tan\tau=(2
\alpha_{v}-1)/(2 \alpha_{u}-1)$, and $\lambda$ is a tunable
parameter. Here $\alpha_{u,v}$ are the relative frequencies of the
outcomes as defined in section~\ref{fixed local}. Note that in
average $\omega$ will be small since we expect
$\alpha_{u,v}\approx 1/2$, and only terms up to order $\omega^2$
need to be retained. The fidelity can be computed from~(\ref{f-1})
yielding
\begin{equation}
    F\gtrapprox1-(1-\la)^2(1-F_{0})-\la^2{1-4(1-F_0)\over N-N_{0}}+\cdots,
    \label{F-F0}
\end{equation}
where $F_{0}$ is the optimal fidelity for the first $N_{0}$
measurements and the dots stand for subleading terms in inverse
powers of $N$ and $N_{0}$. If $N_0=N^{\beta}$ with
$0<\beta<1$,\footnote{In fact, it can be shown that $\beta=1/2$ is
the best choice for the partitioning of $N$. \cite{us-prep}} it is
clear that the optimal choice is $\la=1$, and then
$F\approx1-1/N$. Therefore local measurements saturate the CM
bound at leading order.

\section{Conclusions}
We have obtained the optimal estimation of the a pure qubit state
for any given set of measurements and and any a priori probability
distribution. We have focussed on local measurement schemes. For
states that are known to lay on the equator of the Bloch sphere
(2D case), we have explicitly shown that, rather surprisingly, the
most basic scheme (local and without classical communication)
saturates the CM bound. This does not happen in the 3D case,
although the basic scheme yields a fidelity very close to the CM
bound. We have also obtained the optimal local scheme and shown
that indeed the CM bound is saturated. Furthermore, numerical
analysis reveals that the CM regime is reached for values of $N$
as low as $12$. Our main conclusion is that CM  do not provide a
significant improvement over local measurements.

Our results can be generalised to other interesting issues, such
as estimating  mixed states,\cite{us-density} unknown unitary
operations, trace preserving maps, etc. For those, the  use of
local measurements is of outmost interest.

\section*{Acknowledgments}
We acknowledge financial support from  MCyT project BFM2002-02588,
CIRIT project SGR-00185 and QUPRODIS EEC contract IST-2001-38877.


\begin{thebibliography}{99}
\bibitem{holevo} A.S.~Holevo, \textit{Probabilistic and Statiscal Aspects
                of Quantum Theory} (North Holland, Amsterdam, 1982).

\bibitem{mp} S.~Massar and S.~Popescu, \textit{Phys. Rev.~Lett.}~{\bf 74}, 1259 (1995).
\bibitem{derka}R.~Derka, V.~Buzek and A.~K.~Ekert, \textit{Phys. Rev. Lett.}~{\bf 80}, 1571 (1998)
\bibitem{pure-states}
            J.~I.~Latorre, P.~Pascual and R.~Tarrach, \textit{Phys.~Rev.~Lett.}~{\bf 81}, 1351 (1998);
            N.~Gisin and S. Popescu,  \textit{Phys.~Rev.~Lett.}~{\bf 83}, 432 (1999);
            E.~Bagan {\em et al}, \textit{Phys.~Rev.~Lett.}~{\bf 85}, 5230 (2000);
                        {\it ibid.} \textit{Phys. Rev.}~A {\bf 63}, 052309 (2001);
            A.~Acin, J.~I.~Latorre and P.~Pascual, \textit{Phys.~Rev.}~A {\bf 61}, 022305(2000);
            A.~Peres and P.~F.~Scudo, \textit{Phys.~Rev.~Lett.}~{\bf 86}, 4160  (2001).
\bibitem{us-product} E.~Bagan, M.~Baig and  R.~Munoz-Tapia,
             \textit{Phys.~Rev.}~A {\bf 64}, 022305(2001).

\bibitem{density}
              J. I. Cirac, A. K. Ekert and C. Macchia\-vello,
              \textit{Phys.~Rev. Lett.}~{\bf 82}, 4344 (1999);
              G.~Vidal {\it et al.}, \textit{Phys. Rev.} A \textbf{60}, 126
              (1999);
              D.~G.~Fischer and M.~Freyberger,
              \textit{Phys.~Lett.} A~\textbf{273}, 293 (2000);
              M.~Keyl and R.~F.~Werner,
              \textit{Phys. Rev.} A~\textbf{64}, 052311 (2001).

\bibitem{locals}K.~R.~Jones, \textit{Phys.~Rev.} A~\textbf{50} 3682 (1994);
           R.~D.~Gill and S.~Massar, \textit{Phys.~Rev.}~A {\bf 61}, 042312 (2000);
           D.~G.~Fisher, S.~H.~Kienle and M.~Freyberger,
           \textit{Phys. Rev.}~A {\bf 61} 032306 (2000);
           Th.~Hannemann \textit{et al}, Phys.~Rev.~A~{\bf 65}, 050303 (2002).

\bibitem{us-local} E. Bagan, M. Baig and R. Munoz-Tapia,
             \textit{Phys.~Rev.~Lett.}~{\bf 89}, 277904 (2002).

\bibitem{us-density}E.~Bagan {\em et al.}, quant-ph/0307199.

\bibitem{tomography}
          A.~G.~White \textit{et al}, \textit{Phys.~Rev.~Lett.}~{\bf 83}, 3102 (1999);
          G.~M~D'Ariano and M.~G.~A.~Paris, \textit{Phys. Rev.}~A {\bf 60} 518 (1999);
          D.~F.~V.~James \textit{et al},  \textit{Phys. Rev.}~A {\bf 64} 052312 (2001);
          G.~M~D'Ariano and P.~Lo~Presti, \textit{Phys.~Rev.~Lett.}~{\bf 86}, 4195 (2001);
          R.~T.~Thew \textit{et al},  \textit{Phys. Rev.}~A {\bf 66} 012303 (2002);
          J.~B~Alepeter \textit{et al},  \textit{Phys.~Rev.~Lett.}~{\bf 90}, 193601 (2003);
          G.~M~D'Ariano, L.~Maccone and M.~Paini, \textit{J.~Opt.}~B {\bf 5}, 77
          (2003).

\bibitem{likelihood}
          Z. Hradil, \textit{Phys. Rev.}~A \textbf{55}, 1561 (1997);
          K.~Banaszek, \textit{Phys. Rev.}~A \textbf{59}, 4797 (1999);
          J~Fiur{\'a}sek and Z~Hradil, \textit{Phys. Rev.}~A \textbf{63}, 020101
          (2001).

\bibitem{us-prep} E.~Bagan {\em et al}, in preparation.



\end{thebibliography}
\end{document}